\documentclass[11pt,twoside,a4paper,cmspaper,final,collab]{cms-tdr}
\def\svnVersion{0b1a7c2-D}\def\svnDate{2026/10/29}\def\cmsCernNoTag{CERN-EP-2026-276}\def\cmsCernDate{\today}\def\cmsMessage{Submitted to Physical Review Letters}
\begin{document}\cmsNoteHeader{HIN-25-013}

\newcommand{\thetac}{\ensuremath{\theta_{\text{c}}\xspace}}
\newcommand{\thetad}{\ensuremath{\theta_{\text{d}}\xspace}}
\newcommand{\thetal}{\ensuremath{\theta_{\text{l}}\xspace}}
\newcommand{\lninvthetal}{\ensuremath{\ln(1/\thetal)}\xspace}

\newcommand{\pp}{\ensuremath{\Pp\Pp}\xspace}
\newcommand{\pTjet}{\ensuremath{\pt^{\text{jet}}}\xspace}

\ifthenelse{\boolean{cms@external}}{\providecommand{\cmsLeft}{upper\xspace}}{\providecommand{\cmsLeft}{left\xspace}}
\ifthenelse{\boolean{cms@external}}{\providecommand{\cmsRight}{lower\xspace}}{\providecommand{\cmsRight}{right\xspace}}
\ifthenelse{\boolean{cms@external}}{\providecommand{\cmsLeftC}{Upper\xspace}}{\providecommand{\cmsLeftC}{Left\xspace}}
\ifthenelse{\boolean{cms@external}}{\providecommand{\cmsRightC}{Lower\xspace}}{\providecommand{\cmsRightC}{Right\xspace}}

\cmsNoteHeader{HIN-25-013}

\title{Probing the charm-quark dead cone effect in heavy ion collisions}

\author[cern]{The CMS Collaboration}
\date{\today}

\abstract{The first experimental study of the dead-cone effect for the charm quark in heavy ion collisions is presented. The analysis is based on a sample of lead-lead (PbPb) collisions at a center-of-mass energy per nucleon pair of 5.02\TeV, recorded by the CMS experiment at the CERN LHC in 2018 and corresponding to an integrated luminosity of 1.7\nbinv. The studied observable is the angular distribution of hard, collinear intrajet emissions that are selected with the late-\kt grooming algorithm. The analysis is performed using inclusive jets and a sample of jets containing prompt \PDz mesons. The angular distribution of late-\kt emissions in PbPb collisions is found to be shifted towards larger angles in \PDz jets relative to inclusive jets, indicating a suppression of collinear splittings for \PDz jets. This shift, which is expected from the dead-cone effect, is consistent with a similar observation made in proton-proton collisions. The new measurement disfavors hypotheses that predict a filling of the charm-quark dead cone by medium-induced radiation in heavy ion collisions.}

\hypersetup{
pdfauthor={CMS Collaboration},
pdftitle={Probing the charm-quark dead cone effect in heavy ion collisions},
pdfsubject={CMS},
pdfkeywords={CMS, charm quark, dead cone, jet substructure, late-kT, soft drop, Lund jet plane, heavy ions, lead-lead}}

\maketitle

Ultrarelativistic heavy ion collisions are studied to understand the properties of the quark-gluon plasma (QGP), a high-temperature, strongly interacting phase of nuclear matter~\cite{Busza:2018rrf,Adams:2012th,Hayrapetyan:2897965}.
The fragmentation of energetic jets, initiated by partons produced in the early stages of the collision, is intertwined with the time evolution of the QGP. 
As a result of parton-QGP interactions, which happen over a wide range of scales,
the jet yields and jet radiation patterns are modified~\cite{Cao:2020wlm}.  
The dominant interaction mechanism between the jet constituents and the QGP, at short distance scales, is expected to be gluon radiation induced by the deconfined medium~\cite{Gyulassy:1990ye,Baier:1996kr,Baier:1996sk}.
The medium is characterized by a ``coherence angle'', \thetac, 
the minimum angle for medium-induced gluons to be decohered and emitted~\cite{Casalderrey-Solana:2012evi}; emissions at smaller angles remain coherent and not resolved as independent color charges by the QGP.

The dead-cone effect in quantum chromodynamics (QCD) is a suppression of gluon emissions off a quark in a cone of angle $\thetad = m_{\PQq}/E_{\PQq}$ around the emitting quark~\cite{Dokshitzer:1991fd}, where $m_{\PQq}$ and $E_{\PQq}$ are the mass and energy of the radiating quark, respectively.
The effect is more pronounced for heavy quarks, being negligible for light quarks.
It was measured in proton-proton (\pp) collisions at the CERN LHC by ALICE~\cite{ALICE:2021aqk} and CMS~\cite{CMS:2025afq} for \PQc quark initiated jets (\PQc jets), and by LHCb~\cite{LHCb:2025mcq} and CMS~\cite{CMS:2025dsz} for \PQb quark initiated jets (\PQb jets), using iterative declustering techniques that give access to the building blocks of the parton shower~\cite{Marzani:2019hun}.
The dead-cone effect implies a reduced number of emissions in heavy-flavor jets, leading to harder fragmentation and fewer intrajet constituents with respect to inclusive jets. 
If medium-induced radiation is also affected by the dead-cone effect, jet quenching should be quark mass dependent. 
The flavor dependence of jet quenching has not yet been studied in detail~\cite{Qin:2015srf}.

In heavy ion collisions, the dead cone was predicted to be filled by QGP-induced gluon radiation~\cite{Armesto:2003jh}. 
More recently, an interplay between the coherence angle of the QGP and the dead-cone angle in heavy-flavor-initiated jets was proposed~\cite{Cunqueiro:2022svx, Andres:2023ymw}, leading to QGP-induced gluons filling the dead cone if $\thetac <  \thetad$.
Otherwise, the emission pattern within the dead cone should remain vacuum-like.
The coherence angle depends on the jet transport coefficient $\hat{q}$, which characterizes the transverse momentum (\pt) broadening per unit path length in the medium, and on the in-medium path length $L$ traversed by the parton, as $\thetac \propto 1/\sqrt{\smash[b]{\hat{q} L^{3}}}$.
The bottom-quark dead cone may overlap with the angular region populated by medium-induced radiation, whereas the smaller charm-quark dead cone may remain approximately vacuum-like.
Jet substructure techniques allow the radiation pattern of a jet to be examined by reconstructing its successive splittings, providing an experimental proxy for the parton shower through the primary Lund jet plane~\cite{Lifson:2020gua,Dreyer:2018nbf}.
By selecting emissions at smaller angles and lower energies within the jet, these techniques enhance the sensitivity to quark-mass effects, enabling the study of the charm-quark dead cone even in high-energy jets~\cite{Cunqueiro:2022svx}.

This Letter presents the first experimental study of the dead-cone effect in heavy ion collisions using jets with a prompt \PDz meson, denoted as ``prompt \PDz jets'' (or ``\PDz jets'').
The angular distribution of jet emissions selected by the late-\kt grooming algorithm~\cite{Cunqueiro:2022svx} is measured for \PDz and inclusive jets.
The distributions are compared with \pp measurements at the same center-of-mass energy~\cite{CMS:2025afq} to probe medium-induced modifications. 

The analysis uses PbPb collisions at a center-of-mass energy per nucleon pair of 5.02\TeV, recorded by CMS in 2018 and corresponding to an integrated luminosity of 1.7\nbinv~\cite{CMS:2025rzq}.
Numerical values are reported in the HEPData record of this analysis~\cite{HEPData}.
The CMS apparatus~\cite{CMS:2008xjf,CMS:2309.05466} is a multipurpose, nearly hermetic detector, designed to trigger on~\cite{CMS:2020cmk,CMS:2016ngn} and identify electrons, muons, photons, and charged and neutral hadrons~\cite{CMS:2020uim,CMS:2018rym,CMS:2014pgm}. 
A global ``particle-flow'' (PF) algorithm~\cite{CMS:2017yfk} aims to reconstruct all individual particles in an event, combining information provided by an all-silicon inner tracker, a crystal electromagnetic calorimeter, and a brass and scintillator hadron calorimeter, operating inside a 3.8\unit{T} superconducting solenoid, with data from gas-ionization muon detectors embedded in the flux-return yoke outside the solenoid. 
Hadron forward calorimeters (HF), made of steel and quartz fibers, extend the pseudorapidity coverage up to $\abs{\eta} \approx 5$.
Events of interest are selected using a two-tiered trigger system~\cite{CMS:2020cmk,CMS:2016ngn}. 
Events with high-\pt jets are collected with an exclusive combination of the minimum bias trigger, requiring coincident signals in the HF calorimeters on both sides of the interaction point, and a trigger that requires a calorimeter object with $\pt > 100$\GeV. 
The collision centrality, expressed as a percentile of the total inelastic hadronic cross section~\cite{Loizides:2017ack}, is evaluated from the transverse energy deposited in both HF calorimeters~\cite{CMS:2011iwn}. 
To focus on medium modification effects, the analysis is restricted to the 30\% most central collisions.

Events generated with \PYTHIA 8.230~\cite{Sjostrand:2014zea}, using the underlying event (UE) tune CP5~\cite{CMS:2019csb}, are embedded in a heavy ion event produced with \HYDJET 1.9~\cite{Lokhtin:2005px}, to include the multiple nucleon-nucleon interactions occurring in a PbPb collision. 
The simulation is tuned to reproduce the global event properties of PbPb collisions, such as the average UE \pt density and the track kinematical distributions. 
To reduce data-to-simulation differences in the jet energy calibration for \PDz jets, the detector-level simulated fragmentation function is reweighted by the ratio of the measured to simulated distributions.
The events are processed through a detailed simulation of the CMS detector, based on the \GEANTfour package~\cite{Agostinelli:2002hh}. 

The jets are clustered from the PF candidates using the anti-\kt algorithm~\cite{Cacciari:2008gp,Cacciari:2011ma} with radius $R = 0.2$.
This small radius is chosen to mitigate the effects of the large UE background,
characteristic of heavy ion collisions, on the jet momentum and substructure~\cite{Andrews:2018jcm}.
The robustness of 
this choice and its sensitivity to the dead-cone effect was reported in Ref.~\cite{Cunqueiro:2022svx}.
The detector-level jets must have $80 < \pTjet < 180$\GeV and $\abs{\eta^{\text{jet}}} < 1.6$, while the fully-corrected results are reported in the 100--120\GeV \pTjet range. 
The wider \pTjet interval is used in the unfolding procedure to account for migrations 
into or out of the final range, 
reducing boundary effects from the 
jet momentum resolution and 
underlying-event fluctuations.

The jet constituents are corrected for the UE contribution using the constituent subtraction method~\cite{Berta:2014eza}.
This approach removes or corrects jet constituents based on the average UE energy density, estimated as described in Ref.~\cite{Kodolova:2007hd}.
Fluctuations of the UE background in the $\eta$--$\phi$ plane smear the \pTjet and its internal structure~\cite{ALICE:2012nbx}.
These migrations are corrected through a two-dimensional unfolding in jet \pt and late-\kt splitting angle space, using a response matrix that maps detector-level to particle-level jets.
Jet energy scale (JES) corrections are applied to account for differences in the jet energy response between data and simulation~\cite{CMS:2011shu,CMS:2016lmd}.
The jet energy resolution (JER) is handled by the unfolding procedure.

The \PDz candidates are reconstructed by combining pairs of oppositely charged tracks with invariant mass within $\pm 0.2$\GeV of the 
\PDz mass~\cite{PDG2026}. 
To suppress the combinatorial background and improve the momentum and mass resolutions, the tracks must have $\pt^{\text{trk}} > 1$\GeV and $\abs{\eta^{\text{trk}}} < 2.4$, and satisfy the CMS high-purity track selection, including requirements on the relative \pt uncertainty, number of tracker hits, and track-fit quality~\cite{CMS:2014pgm}.
The \PDz invariant mass is evaluated for two hypotheses: 
assuming that one track is a pion and the other is a kaon or vice-versa.
The \PDz candidates must have $\pt > 4$\GeV and rapidity $\abs{y} < 1.2$.
The \pt selection reduces the combinatorial background 
in the heavy ion environment~\cite{CMS:2019jis} and 
mitigates residual data-to-simulation differences in the \PDz reconstruction efficiency. 
To distinguish the $\PDz \to \PGpp \PKm$ (and charge conjugate) signal 
from the background, we use a multivariate analysis technique based on a boosted decision tree 
(BDT)~\cite{Roe:2004na}. 
The BDT is trained with the detector-level MC sample for the signal and 
measured events from the mass side-bands for the background, using variables reflecting the topological and kinematical features of the \PDz decay. 
The remaining background is statistically subtracted through fits to the \PDz invariant mass distributions.

The analysis provides the angular distribution of the intrajet emissions, as selected by the late-\kt algorithm, for \PDz jets and inclusive jets.
The \PDz candidates are reconstructed and selected as previously described.
After using the two charged daughter tracks to reconstruct the four-momentum of the \PDz candidate, the daughters are removed from the particle list and replaced by the \PDz, which is treated as a single constituent in the subsequent jet finding and CA reclustering.
The jet substructure observables are then computed for jets containing at least one \PDz candidate.
The constituents of a jet clustered using the anti-\kt algorithm are reclustered using the CA algorithm~\cite{Dreyer:2018nbf}, only selecting PF candidates with $\pt > 1$\GeV to ensure a high tracking efficiency and reduce data-to-simulation differences.
The CA pairwise clustering order is then followed in reverse.
Starting with the reclustered jet $j$, the last step of the CA clustering is undone, so that the original jet is declustered into the two subjets of the previous clustering step, $j \to j_a + j_b$, where $j_a$ ($j_b$) is the harder (softer) subjet in the branching.
The $j_a$ subjet is declustered in the next step, while the softer subjet is treated as an emission. 
The late-\kt groomer selects the last emission in the declustering process whose relative transverse momentum \kt is larger than 1\GeV. 
If the leading prong of the selected splitting contains the \PDz candidate as a constituent, the ``late-\kt splitting angle'', \thetal, defined as the angular distance between the two prongs in the pseudorapidity--azimuthal angle plane, is registered; otherwise the jet is rejected.

\begin{figure}[ht]
\centering
\includegraphics[width=0.46\textwidth]{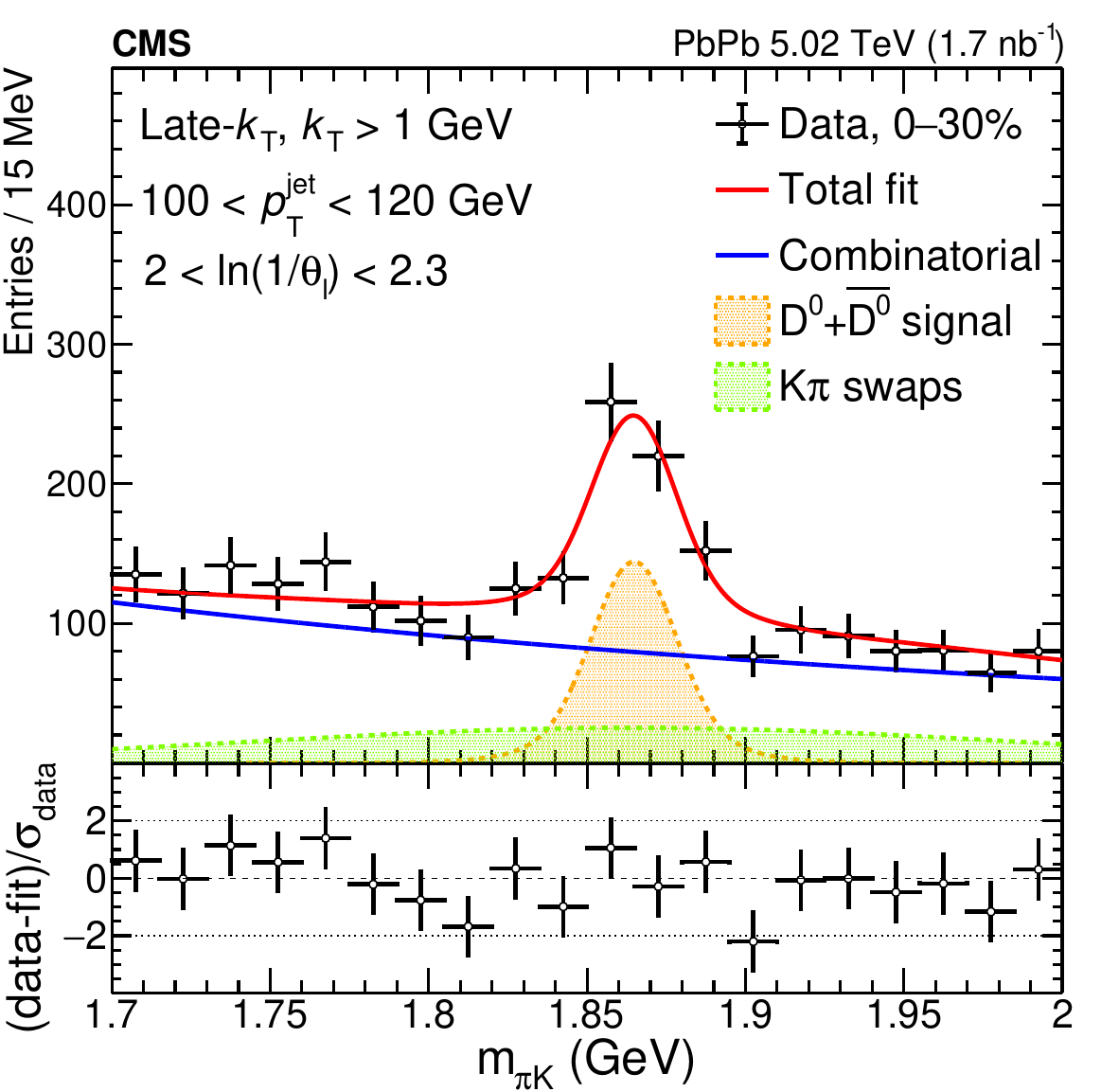}
\caption{The $\PGppm \PKmp$ invariant mass distribution for \PDz jets with $100 < \pTjet < 120$\GeV and $2 < \lninvthetal < 2.3$. 
The fit result and its individual components are shown. 
The vertical bars on the data points represent statistical uncertainties. 
The lower panel shows the difference between the data and the fit result divided by the statistical uncertainty of the data.}
\label{fig:figure1}
\end{figure}

The \PDz signal yield is measured, in each \pTjet and \lninvthetal bin, by fitting the \PDz invariant mass distributions in the 1.7--2.0\GeV range, as illustrated in Fig.~\ref{fig:figure1}. 
The signal shape is modeled by the sum of two Gaussian functions, with common means and different widths constrained by simulation studies, while the  combinatorial background 
is modeled by a power law function. 
An additional Gaussian, with a much larger width also constrained by simulation, models the term resulting from swapped pion and kaon mass assignments.
The $\PDz \to \PKp \PKm$ and $\PDz \to \PGpp \PGpm$ backgrounds (too small to be visible in the figure)
are modeled with a Crystal Ball function~\cite{Skwarnicki:1986xj} each.

\begin{figure}[ht]
\centering
\includegraphics[width=0.45\textwidth]{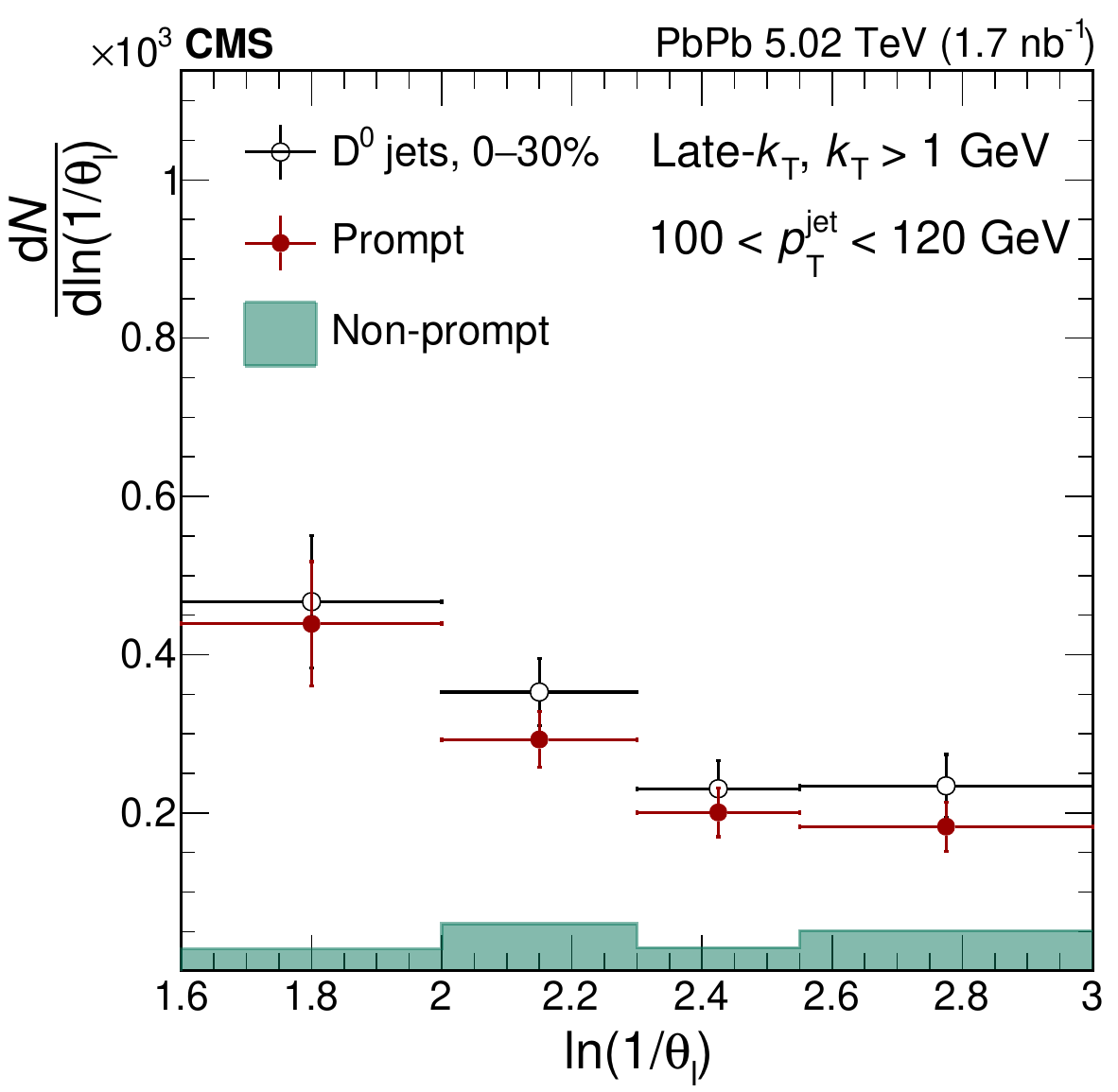}
\caption{The \lninvthetal distributions for \PDz jets (open circles), prompt-\PDz jets (filled circles), and nonprompt \PDz jets (histogram).
The vertical bars represent statistical uncertainties. 
The fraction of jets not satisfying the late-\kt grooming condition is reported in the legend.}
\label{fig:figure2}
\end{figure}

The \PDz yield extracted from fits of the invariant mass distribution includes both a prompt term, from the fragmentation of a charm quark, and a nonprompt term, from decays of \PQb hadrons. 
In the latter case, the reconstructed jet contains a \PDz but the fragmentation and dead-cone effects are driven by the \PQb quark mass.
In addition, the \PQb hadron decay products will also be jet constituents.
The prompt \PDz signal, which includes feed-down contributions from heavier charm hadrons, is obtained using template fits to the distance-of-closest-approach (DCA) significance distribution~\cite{CMS:2022vfn}.
Figure~\ref{fig:figure2} shows the \PDz jet substructure distribution, as a function of the splitting angle, obtained from fits of the invariant mass distributions in bins of DCA and \pTjet. 
The nonprompt contribution, also shown, is approximately 10\%.

The unfolding recovers the particle-level distributions by correcting for detector resolution and underlying-event fluctuations that migrate jets between bins of \pTjet and \lninvthetal.
It also corrects for UE effects, quantified using a four-dimensional response matrix built from the particle- and detector-level simulated distributions.
The particle level is defined by the set of generated particles with lifetimes longer than 10\unit{ns}.
If the particle- and detector-level jets satisfy the late-\kt grooming criteria, the splittings selected at both levels are required to be, in addition, the closest 
among all the primary emissions in the two CA trees. 
This requirement reduces off-diagonal elements in the response matrix arising from uncorrelated emissions, but requires matching-purity and matching-efficiency corrections.
The matching-purity correction removes detector-level splittings without a particle-level match before unfolding; 
it is approximately 45\% at small angles and decreases to 25\% at large angles because of unpaired or misidentified emissions from UE particles.
The matching-efficiency correction recovers particle-level splittings without a detector-level match after unfolding, accounting for detector effects that distort the jet tree; it is approximately 50\%, roughly independent of the splitting angle.
The response matrix describes $(\pTjet,\lninvthetal)$ migrations only for matched splittings.
This factorization reduces migrations between unrelated splittings and improves the stability and diagonality of the response matrix. 
The PbPb corrections are a few percent lower than the \pp values. 
The dependence of the matching purity and efficiency on the modeling of the jet shower is included in the physics-model systematic uncertainty, 
evaluated by recomputing the full correction chain using an alternative particle-level distribution, based on the \textsc{Hybrid} model discussed below.
We use the D'Agostini iterative unfolding with early stopping~\cite{DAgostini:1994fjx}, as implemented in the \textsc{RooUnfold} 2.0.0 package~\cite{RooUnfold}. 
The number of iterations, which serves as the regularization parameter, 
is obtained by comparing the folded-back distribution with the measured input 
and is determined by $\chi^2$ goodness-of-fit tests.
The fraction of generator-level \PDz mesons in each \pTjet bin with both daughters, at the detector level, within the acceptance window and the \PDz reconstruction efficiency, are evaluated with the simulated samples and have values around 90\% and 85\%, respectively. 
The efficiency of the BDT selection is approximately 90\%, independently of the substructure observable under study, confirming that the \PDz selection is based on decay properties uncorrelated with the jet substructure.
All three factors are computed in \pTjet and \lninvthetal bins.
Their product is applied to the unfolded results.

The measurement uncertainties, described in the following, are propagated by repeating the unfolding procedure with variations of the response matrix, of the prior distribution, and of the purity and efficiency corrections. 
The JES and JER corrections have uncertainties of 2--4\%~\cite{CMS:2016lmd}. 
They are propagated through the unfolding by constructing alternative response matrices and lead to a maximum uncertainty of 6\%, at small angles.
To account for the uncertainty in the calibration of each PF object, the four-momenta of the jet constituents are independently shifted by different amounts~\cite{CMS:2017yfk, CMS:2016lmd, CMS-PAS-JME-10-008}, resulting in an alternative response matrix. 
The associated uncertainty is at the level of 2\% for inclusive jets and spans in 2--10\% for \PDz jets. 
The uncertainty in the modeling of the tracking efficiency~\cite{CMS:2016lmd}, which affects both the \PDz reconstruction and the substructure observable, is below 1\% for all angles.
To account for this uncertainty, 4\% of the tracks are randomly removed from the simulated sample prior to the \PDz reconstruction.
At the level of the jet substructure, the lost track is assumed to be a neutral hadron and its energy is smeared by 10\%, reflecting the hadron calorimeter resolution. 

The choice of the physics model impacts the unfolding corrections of the measured observable via the description of the jet showering pattern, which changes the detector response at the subjet level, as well as the prior spectrum used for regularization; it also affects the matching efficiency and purity corrections. 
To assess this model dependence, the \PYTHIA events are reweighted, in the two-dimensional particle-level $(\pTjet,\lninvthetal)$ plane, to match the distribution of the \textsc{Hybrid} model, which includes jet-quenching~\cite{Casalderrey-Solana:2014bpa,Casalderrey-Solana:2020rsj}. 
The associated uncertainty is 1--4\% for \PDz jets and 2--7\% for inclusive jets.

The statistical uncertainties of the simulated sample used to compute the response matrix are propagated through the unfolding procedure, giving a contribution to the covariance matrix.
The \PDz signal extraction uncertainties are evaluated by replacing the double-Gaussian signal and power law background models with
triple-Gaussian and exponential functions, yielding uncertainties of 0.5--8\% and 2--6\%, respectively. 
The systematic uncertainty associated with the prompt \PDz fraction is evaluated by varying that fraction within its uncertainty, as obtained from the fit of the DCA significance distribution, in each $(\pTjet, \lninvthetal)$ bin. 
The variations in the corrected distributions are symmetrized and taken as uncertainty, ranging from 2 to 16.6\%.
The total systematic uncertainties range from 8.6--21.5\% for \PDz jets and 3.3--7.8\% for inclusive jets, dominated by the prompt \PDz fraction, signal yield extraction, and PF scale for \PDz jets, and the physics
model for inclusive jets.

The statistical covariance matrix is propagated through the unfolding and normalization procedures. 
The unfolding introduces nonzero off-diagonal terms that encode bin-to-bin migrations between \pTjet and \lninvthetal, while the subsequent normalization of the angular distribution induces additional bin-to-bin anticorrelations.
These correlations are kept in the statistical uncertainties and corresponding compatibility tests.
The statistical uncertainties are 14.3--16\% for \PDz jets and 0.7--2\% for inclusive jets.
In the \PDz-to-inclusive ratio, we consider as fully correlated the uncertainties related to the tracking efficiency, the PF calibration, and the JES and JER components common to \PDz jets and inclusive jets.
In the PbPb to \pp double ratio, no correlations are considered, as the samples are from different years.

\begin{figure}[ht]
\centering
\includegraphics[width=0.5\textwidth]{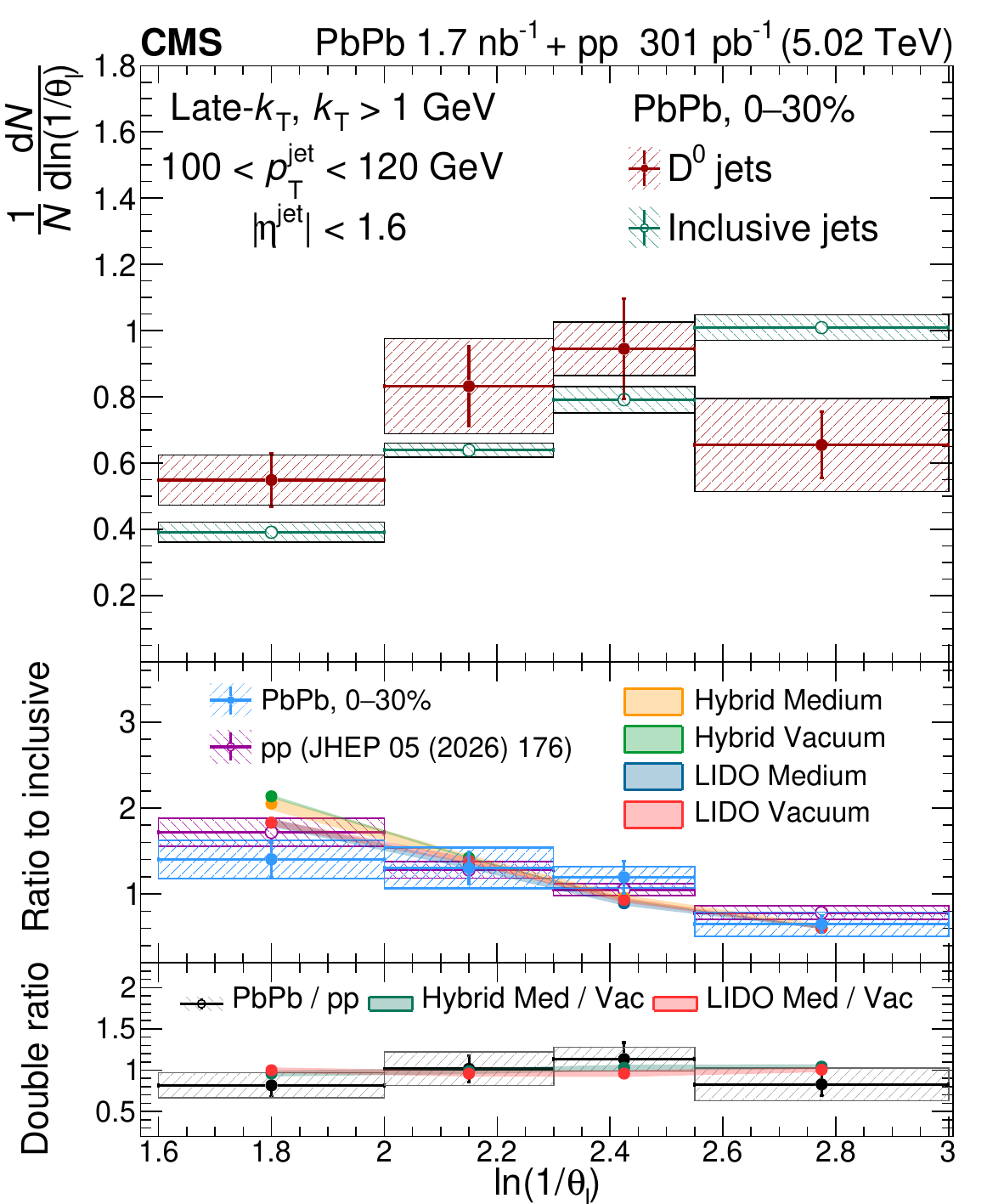}
\caption{Upper: Late-\kt angular distribution for prompt \PDz-tagged (red) and inclusive (cyan) jets. 
Middle: The \PDz-to-inclusive ratio in PbPb (blue) and pp (purple) collisions, 
together with predictions from the \textsc{Hybrid} and \textsc{Lido} quenched models, in the ``medium'' and ``vacuum'' scenarios.
Lower: The PbPb over \pp double ratio.
The vertical bars (boxes) represent statistical (systematic) uncertainties, while the bands represent the theoretical uncertainties.}
\label{fig:finalLatekt}
\end{figure}

Figure~\ref{fig:finalLatekt} presents the fully corrected \PDz jet and inclusive jet \lninvthetal distributions in PbPb collisions (upper panel), their ratio (middle panel), and their double ratio relative to \pp collisions~\cite{CMS:2025afq} (lower panel).
The \PDz jet angular distribution shows a trend towards larger splitting angles (smaller \lninvthetal values) relative to the inclusive jet distribution, reflecting the convolution of the energy distribution of the selected splittings and their individual dead-cone angles.
A similar trend was seen in \pp collisions~\cite{CMS:2025afq} and attributed to the relatively high charm quark mass.
The compatibility of the \PDz-to-inclusive ratio with unity is quantified using a $\chi^2$ test over the four angular bins common to the PbPb and \pp measurements.
We use the full statistical covariance matrix, including bin-to-bin correlations induced by the unfolding and normalization procedures; systematic uncertainties are included as diagonal contributions. The test gives a $p$-value of 0.16. 
Thus, while the measured ratio exhibits the angular trend described above, the hypothesis of identical \PDz jet and inclusive jet angular shapes is not excluded with the present precision.
Studies based on simulated and control data samples show that the measurements are not affected by a fragmentation bias potentially introduced by the 4\GeV \PDz \pt threshold, a small value in comparison with the typical \PDz \pt of around 50\GeV for the reported 100--120\GeV \pTjet range.
In heavy ion collisions, broad jets are known to lose more energy than narrow ones, leading to a selection bias effect~\cite{CMS:2024zjn}: in a given jet momentum window, the selected jets are narrower in PbPb than in \pp collisions.
However, the energy loss of charged particles and of \PDz mesons has been seen to be compatible~\cite{CMS:2017qjw}, so that this selection bias should not influence our measurement. 
The PbPb-to-\pp double ratio is compatible with unity. 
We have also tested the hypothesis that this double ratio is constant and equal to unity over the four angular bins, retaining the available PbPb statistical bin-to-bin covariance and treating the \pp uncertainties as diagonal; the result is $\chi^2=1.55$ for the four tested bins, corresponding to a $p$-value of 0.82.
Since the bin-to-bin covariance information is not available for the published \pp reference measurement, the test uses the total bin-by-bin uncertainties, treating the PbPb and \pp measurements as independent.

The lower panels of Fig.~\ref{fig:finalLatekt} compare the measurements with \textsc{Hybrid}~\cite{Beraudo:2025nvq} and \textsc{Lido}~\cite{Ke:2018tsh} models, in their ``vacuum'' and ``medium'' scenarios.
The vacuum parton-shower baselines are \PYTHIA for both models, with next-to-leading order corrections~\cite{Cacciari:1998it} included in the \textsc{Lido} model. 
The two calculations rely on substantially different descriptions of the jet-medium interaction.
The \textsc{Hybrid} model includes energy loss and parton broadening from interactions with a strongly coupled QGP, as well as heavy-quark medium response and diffusion effects, whereas the \textsc{Lido} model describes the transport of partons in a hydrodynamically evolving QGP through perturbative Boltzmann transport, including collisional and radiative energy loss and the Landau--Pomeranchuk--Migdal interference effect~\cite{Ke:2018jem}.
Despite these differences, both models predict a similar hierarchy between prompt-\PDz-tagged and inclusive jets and a PbPb-to-\pp double ratio close to unity.
In the \textsc{Hybrid} calculation, the medium and vacuum predictions are nearly identical for this observable, whereas in \textsc{Lido} the medium-induced radiation produces a small shift towards larger angles that is similar for inclusive and prompt-\PDz-tagged jets and therefore largely cancels in the ratio.
The similarity of the final predictions suggests that, within the present observable and kinematic range, the flavor dependence is relatively robust against the different implementations of jet quenching considered here, while the measurement has limited discriminating power between the two frameworks.
In the kinematic regime considered here, the charm-quark dead-cone angle is expected to be smaller than the medium coherence angle, $\thetad < \thetac$, suggesting that the medium induced radiation may occur predominantly outside the charm dead-cone region. The present results are consistent with this expectation and with theoretical scenarios in which the charm-quark dead-cone region remains approximately vacuum-like.

In summary, this Letter presents the first characterization of the dead-cone effect for charm quarks in heavy ion collisions.
Novel jet substructure techniques are used to examine potential modifications for charged quark jets at high energies.
The angular distribution of hard, collinear emissions selected by the late-\kt grooming algorithm is measured for inclusive and prompt \PDz-tagged jets. 
Compared to inclusive jets, \PDz jets exhibit a suppression of small-angle emissions, consistent with the expected dead-cone effect. 
Within the probed phase space and experimental precision, no significant differences are seen between the prompt-\PDz over inclusive angular distribution ratios measured in \pp and PbPb collisions. 
These results support that, in PbPb collisions, the charm-quark dead-cone region remains approximately vacuum-like.

\begin{acknowledgments}
We congratulate our colleagues in the CERN accelerator departments for the excellent performance of the LHC and thank the technical and administrative staffs at CERN and at other CMS institutes for their contributions to the success of the CMS effort. In addition, we gratefully acknowledge the computing centres and personnel of the Worldwide LHC Computing Grid and other centres for delivering so effectively the computing infrastructure essential to our analyses. Finally, we acknowledge the enduring support for the construction and operation of the LHC, the CMS detector, and the supporting computing infrastructure provided by the following funding agencies: SC (Armenia), BMFWF and FWF (Austria); FNRS and FWO (Belgium); CNPq, CAPES, FAPERJ, FAPERGS, and FAPESP (Brazil); MES and BNSF (Bulgaria); CERN; CAS, MoST, and NSFC (China); MINCIENCIAS (Colombia); MSES and CSF (Croatia); RIF (Cyprus); SENESCYT (Ecuador); ERC PRG and PSG, TARISTU24-TK10 and MoER TK202 (Estonia); Academy of Finland, MEC, and HIP (Finland); CEA and CNRS/IN2P3 (France); SRNSF (Georgia); BMFTR, DFG, and HGF (Germany); GSRI (Greece); MATE and NKFIH (Hungary); DAE and DST (India); IPM (Iran); SFI (Ireland); INFN (Italy); MSIT and NRF (Republic of Korea); MES (Latvia); LMTLT (Lithuania); MOE and UM (Malaysia); BUAP, CINVESTAV, CONACYT, LNS, SEP, and UASLP-FAI (Mexico); MOS (Montenegro); MBIE (New Zealand); PAEC (Pakistan); MSHE, NSC, and NAWA (Poland); FCT (Portugal); MESTD (Serbia); MICIU/AEI and PCTI (Spain); MOSTR (Sri Lanka); Swiss Funding Agencies (Switzerland); NSTC (Taiwan); MHESI (Thailand); TUBITAK and TENMAK (T\"{u}rkiye); NASU (Ukraine); STFC (United Kingdom); DOE and NSF (USA).
\end{acknowledgments}\section*{Data availability} Release and preservation of data used by the CMS Collaboration as the basis for publications is guided by the  \href{https://doi.org/10.7483/OPENDATA.CMS.1BNU.8V1W}{CMS data preservation, re-use and open access policy}.

\bibliography{auto_generated}
\cleardoublepage \appendix\section{The CMS Collaboration \label{app:collab}}\begin{sloppypar}\hyphenpenalty=5000\widowpenalty=500\clubpenalty=5000\cmsinstitute{Yerevan Physics Institute, Yerevan, Armenia}
{\tolerance=6000
A.~Belyaev\cmsorcid{0000-0003-1692-1173}, A.~Gevorgyan\cmsorcid{0000-0003-2751-9489}, A.~Hayrapetyan, A.~Tumasyan\cmsAuthorMark{1}\cmsorcid{0009-0000-0684-6742}
\par}
\cmsinstitute{Marietta Blau Institute for Particle Physics, Vienna, Austria}
{\tolerance=6000
W.~Adam\cmsorcid{0000-0001-9099-4341}, L.~Benato\cmsorcid{0000-0001-5135-7489}, T.~Bergauer\cmsorcid{0000-0002-5786-0293}, M.~Dragicevic\cmsorcid{0000-0003-1967-6783}, S.~Gundacker\cmsorcid{0000-0003-2087-3266}, A.K.~Guven\cmsorcid{0009-0004-5670-5138}, M.~Jeitler\cmsAuthorMark{2}\cmsorcid{0000-0002-5141-9560}, N.~Krammer\cmsorcid{0000-0002-0548-0985}, A.~Li\cmsorcid{0000-0002-4547-116X}, D.~Liko\cmsorcid{0000-0002-3380-473X}, M.~Matthewman, A.~Pfeiffer\cmsorcid{0000-0001-5328-448X}, J.~Schieck\cmsAuthorMark{2}\cmsorcid{0000-0002-1058-8093}, R.~Sch\"{o}fbeck\cmsAuthorMark{2}\cmsorcid{0000-0002-2332-8784}, M.~Shooshtari\cmsorcid{0009-0004-8882-4887}, N.~Van~Den~Bossche\cmsorcid{0000-0003-2973-4991}, W.~Waltenberger\cmsorcid{0000-0002-6215-7228}, C.E.~Wulz\cmsAuthorMark{2}\cmsorcid{0000-0001-9226-5812}
% ...
\par}
$^{1}$Also at Yerevan State University, Yerevan, Armenia\\
$^{2}$Also at Technische Universit\"{a}t Wien, Vienna, Austria\\
% ...
\end{sloppypar}
\end{document}